\documentclass[aps,12pt,superscriptaddress,amsfonts,amssymb,amsmath]{revtex4}

\usepackage{graphicx}
\usepackage{epsfig}
\usepackage{makeidx}
\usepackage{epstopdf}

\begin{document}

\title{Time in quantum mechanics and the local \\ non-conservation of the probability current}

\author{G.\ Modanese \footnote{Email address: giovanni.modanese@unibz.it}}
\affiliation{Free University of Bolzano-Bozen \\ Faculty of Science and Technology \\ I-39100 Bolzano, Italy}

\linespread{0.9}

\begin{abstract}

\bigskip

In relativistic quantum field theory with local interactions, charge is locally conserved. This implies local conservation of probability for the Dirac and Klein-Gordon wavefunctions, as special cases; and then in turn for non-relativistic quantum field theory and for the Schr\"odinger and Ginzburg-Landau equations, regarded as low energy limits.
Quantum mechanics, however, is wider than quantum field theory, as an effective model of reality. For instance, fractional quantum mechanics and Schr\"odinger equations with non-local terms have been successfully employed in several applications. The non-locality of these formalisms is strictly related to the problem of time in quantum mechanics.
We compute explicitly for continuum wave packets the terms of the fractional Schr\"odinger equation and of the non-local Schr\"odinger equation by Lenzi et al.\ which break the local current conservation, and discuss their physical significance. The results are especially relevant for the electromagnetic coupling of these wavefunctions.
A connection with the non-local Gorkov equation for superconductors and their proximity effect is also outlined.

\end{abstract}

\maketitle

\section{Introduction}
\label{introduction}

In most physical models, including those of elementary particles and condensed matter, the continuity equation $\partial_t \rho + \nabla \cdot {\bf j}=0$ is of fundamental importance, expressing the basic concept (or we could say the axiom) that the variation in time of the number of particles inside a certain region is opposite to the number of those crossing the boundary of that region in the same time. To visualize the essence of this statement it is also sufficient to think of a 1D system, like a fluid flowing in a pipe or charge carriers flowing in a wire, in which case the equation simply reads $\partial_t \rho+\partial_x j=0$, where $j=\rho v$. For a stationary flow ($\partial_t \rho=0$) this further simplifies, prescribing that $\rho v$ must be independent from $x$.

We know that in quantum mechanics classical concepts like particle trajectory, instant position and velocity, non-invasive measurement, arrival time etc.\ lose their familiar meaning. Nevertheless, the continuity equation is recovered in a probabilistic sense, and one can usually suppose to resort to some classical limit in order to make contact with the corresponding classical concept. In other words, one can still imagine that the ``bean counting'' implied by the continuity equation is still meaningful, albeit in a statistical sense. This is actually true for the Schr\"odinger equation, the Dirac equation and their extensions in quantum field theory. For example, for the Schr\"odinger equation $\ i\hbar \partial_t \Psi=H\Psi \ $ the continuity equation 
takes the form
\begin{align}
	\partial_t \rho + \nabla  \cdot {\bf{j}} = 0; \qquad \rho  = |\Psi {|^2}; \qquad {\bf{j}} = \frac{{ - i\hbar }}{{2m}}\left( {{\Psi ^*}\nabla \Psi  - \Psi \nabla {\Psi ^*}} \right) .
	\label{cont}
\end{align}

Conversely, we can regard the Schr\"odinger equation as a low energy limit of a Dirac or Klein-Gordon equation, and these in turn as consequences of a relativistic quantum field theory with local interactions, where the local conservation of charge is mathematically implemented at a fundamental level, in the definition of the Lagrangian. Also note (and this is not a casual coincidence in our opinion) that in such a theory space and time are essentially treated as two quantities of the same nature. The most radical expression of this equivalence between space and time is Euclidean quantum field theory, which gives the same results for physical observables as the Lorentzian theory \cite{cheng1984gauge,field2001space}. In general, the formalism of elementary particle physics has a strong reductionist character, such that the concept of individual particles, even if weakened in a probabilistic sense, does not allow to conceive exceptions to the law of local charge conservation. In the applications of quantum field theory to condensed matter this attitude is less pronounced, but still present \cite{fetter2012quantum}.

However, when we detach ourselves a little bit from the paradigm of interaction between individual particles, some problems typical of quantum mechanics appear. One of these is the problem of the definition, calculation and measurement of the tunnelling time of a particle across a barrier. This motion does not have a classical analogue, so a correspondence limit of the continuity equation is not applicable. The tunnelling time problem \cite{landauer1994barrier,davies2005quantum} and more generally the problem of arrival time in quantum mechanics \cite{muga2000arrival} have been early recognized and widely studied. Behind these issues lies the fundamental problem of the definition of a time operator in quantum mechanics \cite{delgado1997arrival,toller1999localization}.

The unification of quantum mechanics and special relativity attained in quantum field theory, though successful, inevitably requires a limitation of the scope of quantum mechanics \cite{peres2004quantum}. But in fact some useful quantum mechanical models exist, which are explicitly non-local. One of these 
is the fractional Schr\"odinger equation \cite{laskin2002fractional,Lenzi2008fractional}, where the kinetic energy operator is proportional to $|\textbf{p}|^\alpha$, $1<\alpha \le 2$, leading to a considerable change in the behavior of the wave function. Wei \cite{wei2016comment} has recently shown that its continuity equation contains an anomalous source term, 
taking the form $\partial_t + \nabla  \cdot {\bf{j}} = I_\alpha$, where $I_\alpha$ is called ``extra-current''. This property is related the non-locality of the fractional Schr\"odinger equation \cite{jeng2010nonlocality}. We shall discuss it in Sect.\ \ref{wav}.

The fractional Schr\"odinger equation has recently found novel important applications in optics, in particular concerning the behavior of wavepackets in a harmonic potential \cite{zhang2015propagation}. The theory has further been extended to equations with a periodic $\textsl{P-T}$ symmetric potential; this suggests that a real physical system, the honeycomb lattice, is a possible realization of the fractional Schr\"odinger equation (\cite{zhang2017unveiling} and refs.).
Resonant mode conversions and Rabi oscillations in Gaussian and periodic potentials have been studied in \cite{zhang2017resonant}. While in our present calculations only free wave packets are considered, results in this field have potential applications in diffraction-free and self-healing optoelectronic devices.

Another well-known non-local wave equation of the Schr\"odinger kind is the equation
\begin{align}
	i\hbar\frac{\partial}{\partial t}\Psi(x,t)=-\frac{\hbar^2}{2m} \frac{\partial^2}{\partial x^2} \Psi(x,t) + \int_0^t d\tau \int_{-\infty}^{+\infty} dy \, U(x-y,t-\tau) \Psi(y,\tau).
	\label{lenzi1}
\end{align}
It contains the non-local interaction kernel $U$. Its solutions have been studied in \cite{Lenzi2008solutions,sandev2014time,sandev2016effective}. Possible applications include dissipative quantum
transport processes in quantum dots \cite{baraff1998model,ferry1999complex}. We shall analyze eq.\ (\ref{lenzi1}) and compute the extra-current in Sect.\ \ref{lenzi}. Further applications concern long range interactions \cite{latora1999superdiffusion}, diffusion in active intracellular transport \cite{caspi2000enhanced} and nuclear scattering \cite{chamon1997nonlocal,balantekin1998green}. 

The Gorkov equation for the superconducting order parameter $F (\textbf{x})$ also has a non-local form, namely  
\begin{align}
	F (\textbf{x})=\int d^3 y \, K(\textbf{x},\textbf{y}) F (\textbf{y}).
	\label{eqgorkov}
\end{align}
In Sect.\ \ref{gorkov} we shall se that this equation reduces to a local Ginzburg-Landau equation only under certain assumptions. {It is not known yet whether the full non-local equation leads to an extra-current, also because the form of the kernel $K$ varies considerably, depending on several microscopic assumptions. Nevertheless, this equation deserves further analysis because it can describe weak links in superconductors and Josephson junctions, which in our opinion are good candidates for an experimental observation of the non-local electromagnetic coupling (Sect.\ \ref{emc}); this will be the subject of a forthcoming work.}

In calculations of transport properties of nano-devices such as the density functional theory and non-equilibrium Green function theory, non-local potentials appear in several cases (\cite{li2008definition} and refs.).

{More generally, non-local field theory is a wide subject. Non-local models can originate from coarse graining effective approximations of local models, or they can be conceived even at a more fundamental level than local theories. See on this point the work \cite{kegeles2016generalized}, on which we report further in our Conclusions. In this paper we take a pragmatic stance about non-local wave equations. We do not require them to derive from an action principle or from a more fundamental underlying theory; we regard them just as an effective phenomenological description of certain systems. After all, quantum mechanics was born in this way and has always resisted ``more fundamental'' interpretations.
}

In Sect.\ \ref{emc} we shall discuss the electromagnetic coupling of these wavefunctions, which requires an extension of Maxwell theory. Such extension turns out to satisfy two crucial requirements: (1) It is fully compatible with special relativity. (2) It has a censorship property, according to which the electromagnetic field of a source that is not locally conserved is the same as the field of a fictitious, equivalent (but different) locally conserved source. Thanks to these two properties, the non-local ``quantum anomalies'' that we have demonstrated do not spill over to classical electromagnetic fields, which thus remain a bedrock of special relativity. 

Sect.\ \ref{conc} contains our conclusions.

\section{Wave packet in fractional quantum mechanics}
\label{wav}

Classical fractional diffusion equations have been an active field of research for a long time \cite{evangelista2018fractional,dubkov2008levy,metzler2000random}. Fractional quantum mechanics was started in 2002 with the work by N.\ Laskin on the fractional Schr\"odinger equation \cite{laskin2002fractional} 
\begin{equation}
i\hbar \frac{\partial \Psi}{\partial t} =
\frac{{\hat{\bf p}}^\alpha}{2m} \Psi + V\Psi, \qquad 1<\alpha<2
\end{equation}
and has now several applications, including to statistical physics \cite{Lenzi2008fractional}. 

\subsection{General expression of the current and extra-current}

As pointed out by Wei \cite{wei2016comment}, the probability density $|\Psi|^2$ of the fractional Schr\"odinger equation does not satisfy a continuity equation, but has a non-vanishing ``extra-current'' $I$ defined as
\begin{equation}
I_\alpha = \frac{\partial \rho}{\partial t} + \nabla \cdot {\bf j}=
-i\hbar^{\alpha-1} D_\alpha \left[\nabla \Psi^* (-\nabla^2)^{\frac{\alpha}{2}-1} \nabla \Psi - c.c. \right]
\label{weiI}
\end{equation}
where $D_\alpha$ is a constant with dimension erg$^{1-\alpha}$ cm$^{\alpha}$ s$^{-\alpha}$; when $\alpha=2$ ordinary quantum mechanics is recovered, with $D_2=1/2m$.

 We recall that the current density is given in fractional quantum mechanics by
\begin{align}
	\textbf{j}_\alpha = -i\hbar^{\alpha-1} D_\alpha \left[ \Psi^* (-\nabla^2)^{\frac{\alpha}{2}-1} \nabla \Psi - c.c. \right].
\end{align}

Starting from the definition (\ref{weiI}), in \cite{wei2016comment} $I_\alpha$ was computed for the superposition of two plane waves. In \cite{Modanese2017480} we found $I_{3/2}$ for a non-normalizable discrete packet. In that case $I_{3/2}$ can be written explicitly and a plot was given in Fig.\ 2 of Ref.\ \cite{modanese2017electromagnetic}. It is also possible to plot the time dependence of $|\Psi|$, which shows that the packet moves without spreading (Fig.\ 1 of \cite{modanese2017electromagnetic}). In view of the form of $I_{3/2}$, one can interpret the sub-diffusive movement of the packet as involving a steady backward internal displacement of probability without a corresponding current (there is a net destruction of probability on the right of the maximum and a net creation on the left, when the packet moves to the right).

\begin{figure}
\begin{center}
\includegraphics[width=10cm,height=6cm]{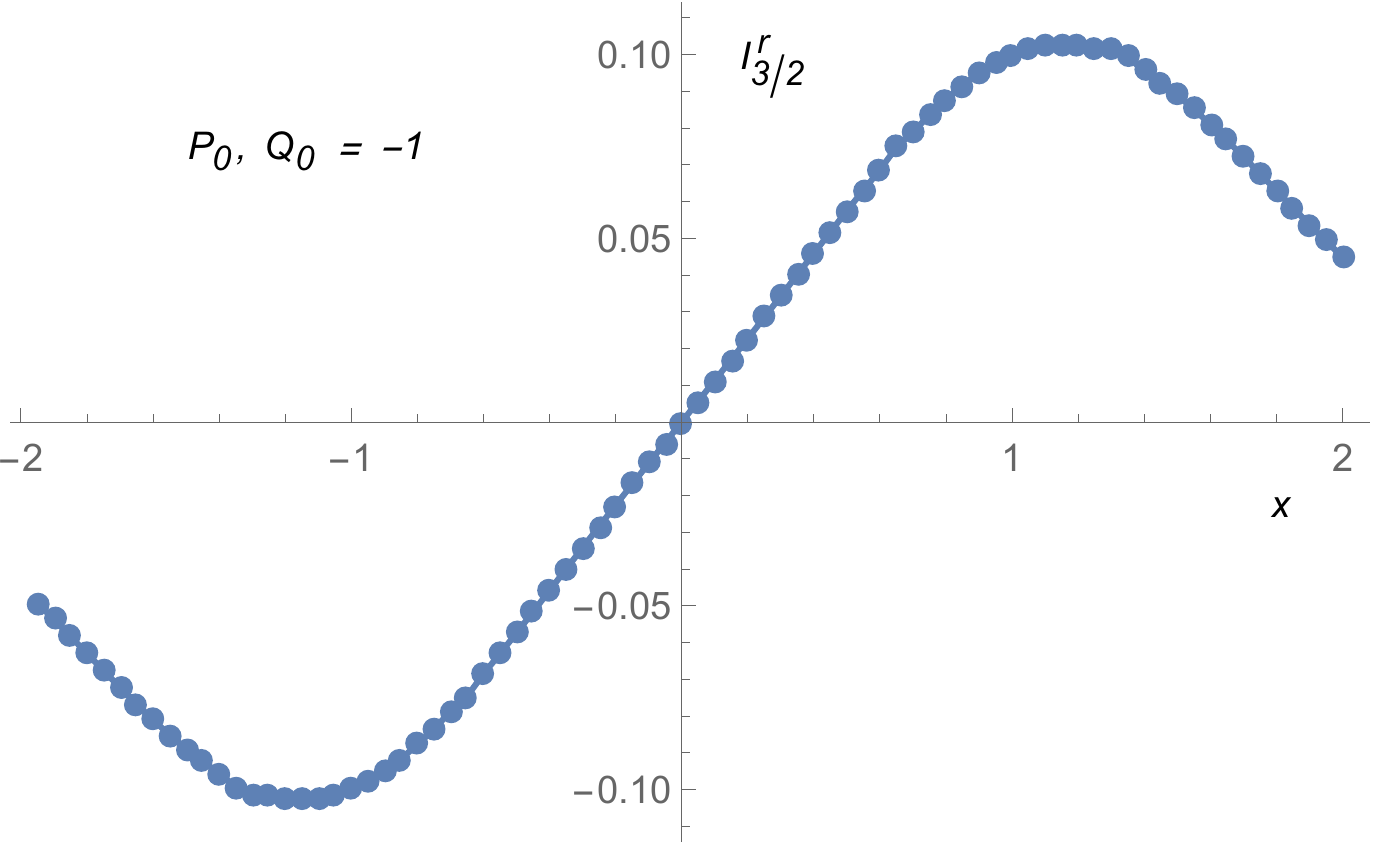}
\caption{Re-scaled extra-current $I^r_{3/2}(X)$ for a wave packet in fractional quantum mechanics with $\alpha=3/2$ that is moving to the left. The extra-current is equal by definition to the quantity $\partial_t \rho + \nabla \cdot {\bf j}$, which vanishes in a theory with local conservation of charge. It can be interpreted in this case as an amount of charge per unit time that disappears from the head of the packet and re-appears in its tail.} 
\label{fig1}
\end{center}  
\end{figure}

\begin{figure}
\begin{center}
\includegraphics[width=10cm,height=6cm]{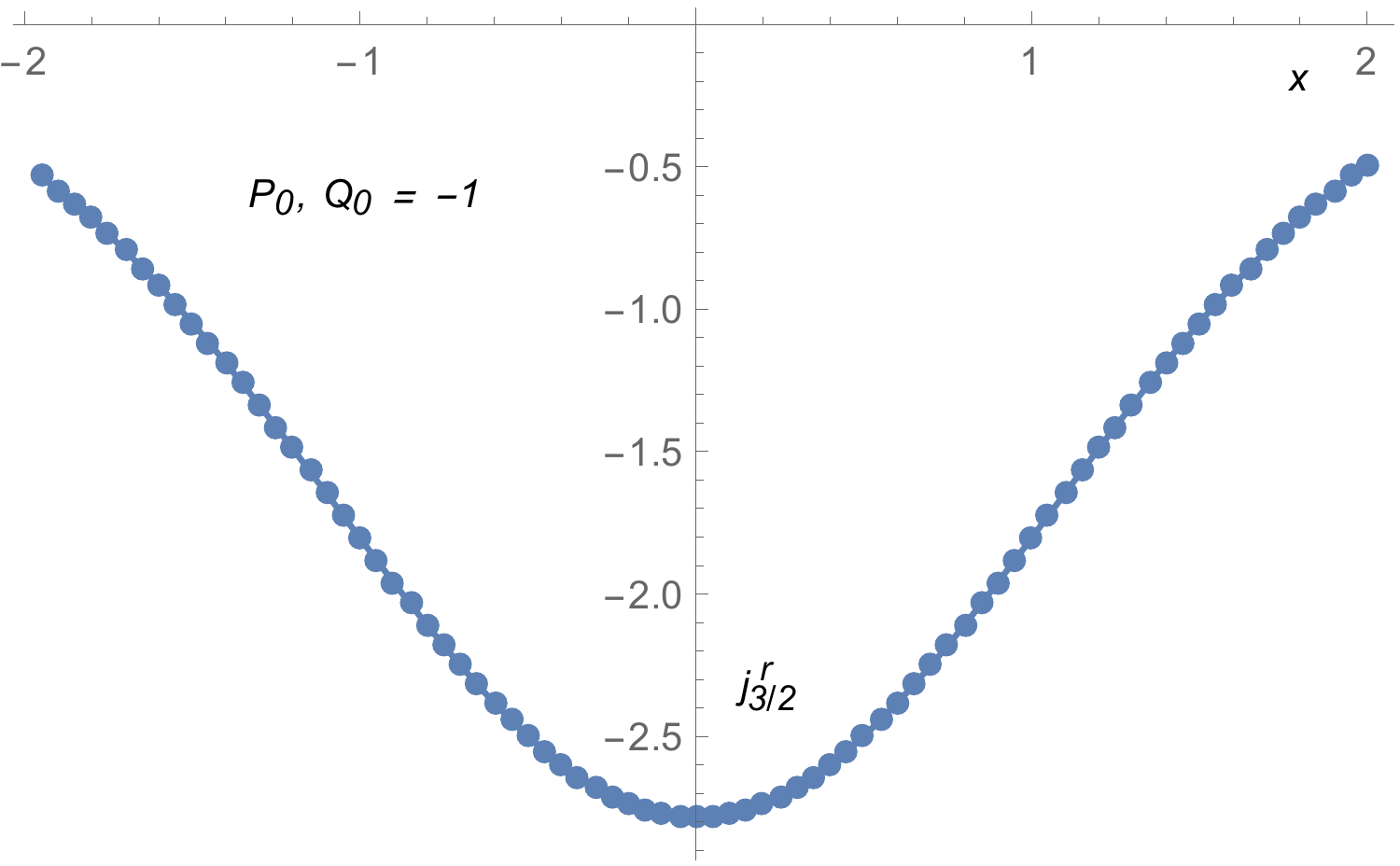}
\caption{Re-scaled current density $j^r_{3/2}(X)$ for the same packet as in Fig.\ \ref{fig1}. It is important to note here (for the comparison of the magnitude orders of $I$ and $j$ in the microscopic case, see Sect.\ \ref{resc}) that the derivative $\partial j / \partial X$ is almost of the same magnitude order as $I$.} 
\label{fig2}
\end{center}  
\end{figure}

Here we show that more general results can be obtained with a continuous packet of the form 
\begin{align}
	\Psi(t,\textbf{x})=\frac{1}{(2\pi)^3}\int d^3p \, e^{i\hbar^{-1}\textbf{p}\textbf{x}} \tilde{\Psi}(t,\textbf{p}).
\end{align}
The operator $(-\nabla^2)^{\frac{\alpha}{2}-1}$ acts on $\tilde{\Psi}$ multiplying it by $|\textbf{p}|^{\alpha-2}$. So we obtain for $I_\alpha$ with some straightforward steps the integral
\begin{align}
	I_\alpha(t,\textbf{x})=-i\hbar^{-1} D_\alpha \int d^3p \, \int d^3q \, \left[ e^{i\hbar^{-1}(\textbf{p}-\textbf{q})\textbf{x}} (\textbf{p}\textbf{q}) |\textbf{p}|^{\alpha-2} \tilde{\Psi}(t,\textbf{p}) \tilde{\Psi}^*(t,\textbf{q}) - c.c. \right].
	\label{intI}
\end{align}
This integral can be computed numerically, with an appropriate ansatz for $\tilde{\Psi}$. In the following we shall consider a Gaussian packet.

We also need an expression for the current density, since in the end we must compare $I_\alpha$ with $\nabla \cdot \textbf{j}_\alpha$, and so indirectly with $\partial_t \rho$. We obtain
\begin{align}
	\textbf{j}_\alpha (\textbf{x}) = -i D_\alpha 
	\int d^3p \, \int d^3q \, \left[ e^{i\hbar^{-1}(\textbf{p}-\textbf{q})\textbf{x}} (i\textbf{p}) |\textbf{p}|^{\alpha-2} \tilde{\Psi}(\textbf{p},t) \tilde{\Psi}^*(\textbf{q},t) - c.c. \right].
	\label{corr}
\end{align}

\subsection{Computation of the re-scaled current and extra-current}

Now we focus on a 1D case, with $\alpha=3/2$, at the time $t=0$. Also in view of the numerical evaluation needed, we first write re-scaled expressions for $I$ and $j$. We choose the physical units in such a way that $\hbar=1$, $D_{3/2}=1$. We call $X$ the re-scaled coordinate and $P$, $Q$ the re-scaled momenta. The integrals for the re-scaled $I$ and $j$ (denoted with a superscript $r$) are the following:
\begin{align}
	I^r_\frac{3}{2}(X)=-i\int dP \int dQ \left[ e^{i(P-Q)X} (PQ) |P|^{-1/2} \tilde{\psi}(P) \tilde{\psi}^*(Q)-c.c. \right];
    \label{eq5}
\end{align}
\begin{align}
	j^r_\frac{3}{2}(X)=-i\int dP \int dQ \left[ e^{i(P-Q)X} (iP) |P|^{-1/2} \tilde{\psi}(P) \tilde{\psi}^*(Q)-c.c. \right].
    \label{eq6}
\end{align}

The re-scaled wavefunctions in momentum space at the initial time are normalized real Gaussians with $\sigma=1$:
\begin{align}
	\tilde{\psi}(P)=\frac{1}{\sqrt{2\pi}} e^{-\frac{1}{2}(P-P_0)^2}.
     \label{eq7}
\end{align}

The double integrals for $I^r_\frac{3}{2}$ and $j^r_\frac{3}{2}$ in eqs.\ (\ref{eq5}), (\ref{eq6}) can be easily computed numerically as functions of $x$. Figs.\ \ref{fig1}, \ref{fig2} show the results for a wave packet $\tilde{\psi}(P)$ of the form (\ref{eq7}), with $P_0=1$, i.e., which is moving to left ($j^r$ negative). This is the complementary case of the discrete packet of Ref.\ \cite{modanese2017electromagnetic} which is moving to the right. The physical interpretation is the same: there is sub-diffusion, in the sense that some probability is subtracted from the head of the traveling packet and transferred instantly to the tail.

The numerical solutions also show that if the packet is at rest, then $I=0$ at $t=0$. This can be proven analytically in a similar way as will be done later for the non-local equation of Lenzi et al., see eqs.\ (\ref{ex24}), (\ref{ex25}). At $t>0$, $I$ is not zero for a packet at rest, and in fact one has anomalous diffusion also in the spreading of the packet, as will be shown in Sect.\ \ref{lenzi}.

\subsection{Ratio between current variations and extra-current on a microscopic scale}
\label{resc}

Looking now at the quantities $\partial_x j$ and $I$ on a microscopic scale, we want to check that their ratio is approximately of magnitude order 1, i.e., the space variation in the flowing current is of the same order as the charge which appears and disappears at different places without a corresponding flow. It could actually happen that, after restoring the dimensional parameters set to unity in the numerical evaluation, one finds $\partial_x j$ to exceed $I$ by several magnitude orders, making the violation of local conservation practically irrelevant. We shall see, however, that this is not the case.

In MKS units, $\hbar \simeq 6\cdot 10^{-34}$. As lenght scale we shall introduce, having in mind atomic and nano-scale systems, the quantity $a=10^{-10}$ m. For the momentum scale we define $b=6\cdot 10^{-24}$ kg$\cdot$m/s, in such a way that $ab=\hbar$. The re-scaled coordinates and momenta are connected to the true microscopic coordinates and momenta through $P=ap$, $X=bx$. The true initial wavefunction in terms of $p$ is 
\begin{align}
	\tilde{\psi}(p)=\frac{1}{a \sqrt{2\pi}} e^{-\frac{1}{2a^2}(p-p_0)^2}
\end{align}
where we have taken the width $\Delta p$ of the wavefunction in momentum space equal to the momentum scale $b$. With this choice, the relation $ab=\hbar$ simply implies that $\Delta p \Delta x=\hbar$ for this wave packet (minimal uncertainty).

When we insert $\tilde{\psi}(p)$ into (\ref{eq5}) and (\ref{eq6}) and change the integration variables as $p \to P/a$, $q \to Q/a$, $x \to X/b$, in the 1D case we obtain for the true (not re-scaled) $j$:
\begin{align}
j_\frac{3}{2}(x)=-i D_\frac{3}{2} \int dP\cdot a \int dQ\cdot a \left[ e^{i(PX-QX) } (iaP) |aP|^{-\frac{1}{2} } \frac{1}{2\pi a^2} e^{-\frac{1}{2} (P-P_0)^2 } e^{-\frac{1}{2} (Q-Q_0)^2 } - c.c. \right].
\end{align}
Similarly we obtain for the true $I$
\begin{align}
I_\frac{3}{2}(x)=-i \frac{D_\frac{3}{2}}{\hbar} \int dP\cdot a \int dQ\cdot a \left[ e^{i(PX-QX) } (a^2QP) |aP|^{-\frac{1}{2} } \frac{1}{2\pi a^2} e^{-\frac{1}{2} (P-P_0)^2 } e^{-\frac{1}{2} (Q-Q_0)^2 } - c.c. \right].
\end{align}
In conclusion we obtain for $j$ and $I$ integrals which coincide with those for $j^r$ and $I^r$, with an additional factor $D_{3/2}a^{\frac{1}{2} }$ for $j$ and $D_{3/2}a^{\frac{3}{2} } \hbar^{-1}$ for $I$.

Now compare $I$ with $\partial_x j$. As magnitude order, $\frac{\partial j}{\partial x}$ is equal to
\begin{align}
\frac{j}{b} \approx D_\frac{3}{2} a^{\frac{1}{2} } \frac{1}{b} = D_\frac{3}{2} a^{\frac{3}{2} } \frac{1}{\hbar},
\end{align}
which is the same as the magnitude order of $I$.

\section{The non-local wave equation by Lenzi et al.\ }
\label{lenzi}

It has the form \cite{Lenzi2008solutions} 
\begin{align}
	i\hbar \frac{\partial}{\partial t} \Psi(x,t) = -\frac{\hbar^2}{2m} \frac{\partial^2}{\partial x^2} \Psi(x,t)+\int_0^t d\tau \int dy U(x-y,t-\tau) \Psi(y,\tau)
	\label{onda-Lenzi}
\end{align}
where $U$ is a non-local interaction kernel.

The continuity equation for the wave equation (\ref{onda-Lenzi}) is found in \cite{Lenzi2008solutions} to be the following, where the argument of $\Psi$, when missing, is understood to be $(x,t)$:
\begin{align}
	\frac{\partial}{\partial t} (\Psi^*\Psi) &= -\frac{i\hbar}{2m} \frac{\partial}{\partial x} \left( \Psi^* \frac{\partial \Psi}{\partial x}-\Psi \frac{\partial \Psi^*}{\partial x} \right) + \\ \nonumber
	&-\frac{i}{\hbar} \int_0^t d\tau \int dy U(x-y,t-\tau) \left[ \Psi(y,\tau)\Psi^*(x,t) - \Psi(x,t) \Psi^*(y,\tau) \right]  .
\end{align}

The term with the double integral is what we have called ``extra-current $I$'' in the case of fractional quantum mechanics. According to Ref.\ \cite{Lenzi2008solutions}, it is possible to prove that if $U(x,t)=\delta(t) \bar{U}(x)$, with $\bar{U}$ a real and symmetric function, then the total probability is conserved. We shall restrict our attention to this case, which also excludes memory effects, due to the presence of $\delta(t)$. We will compute the extra-current using the solution given in \cite{Lenzi2008solutions} for a potential of the form $U(x,t) \propto \delta(t)/|x|^{1+\beta}$, with $1<\beta<2$. The function $U(k,s)$, defined as the Fourier transform in $x$ and Laplace transform in $t$, is $U(k,s)=U_0 |k|^\alpha$, where $U_0$ is a constant.

By applying a Fourier transform to the wave equation (\ref{onda-Lenzi}) Lenzi et al.\ obtain the integral equation
\begin{align}
i\hbar \frac{\partial}{\partial t} \Psi(k,t) = \frac{\hbar^2 k^2}{2m} \Psi(k,t) + \int_0^t d\tau U(k,t-\tau)\Psi(k,t).
\end{align}
Then they apply a Laplace transform and obtain the algebraic equation
\begin{align}
i\hbar  [s\Psi(k,s)-\Psi(k,0)]=\frac{\hbar^2 k^2}{2m} \Psi(k,s)+U(k,s)\Psi(k,s)
\end{align}
where $\Psi(k,0)$ is the Fourier transform of the initial condition; the solution is, with the choice done for $U$,
\begin{align}
\Psi(k,s)=\frac{\Psi(k,0) }{s+i\hbar k^2/(2m) + iU_0 \hbar^{-1}|k|^\beta }.
\end{align} 
With an inverse Laplace transform one arrives to
\begin{align}
\Psi(k,t)=\Psi(k,0) \exp \left[ -\frac{it}{\hbar} \left( \frac{\hbar^2 k^2}{2m} + U_0 |k|^\beta \right) \right] .
\label{A1}
\end{align}
The inverse Fourier transform of this equation can be written as a convolution with the Fox $H$ function. Fig.\ 1 of Ref.\ \cite{Lenzi2008solutions} shows the behavior of $|\Psi(x,t)|$ as a function of $x$ at the times $t=1$, $t=1.5$, $t=2$, with the parameters choice $\beta=3/2$, $\hbar=1$, $m=1$, $U_0=1$, and the initial condition $\Psi(x,0)=e^{-x^2/2}/\pi^{1/4}$.
Our goal, however, is not to compute $\Psi$, but to compute the corresponding
extra-current, given by the formula 
\begin{align}
I(x,t)=-\frac{i}{\hbar}\int_0^t d\tau \int dy U(x-y,t-\tau) [\Psi^*(x,t)\Psi(y,\tau) - c.c.].
\label{extra1}
\end{align}

In the following we turn to units such that $\hbar=1$, $m=1$. In (\ref{extra1}) we express the two wavefunctions and $U$ as inverse transforms:
\begin{align}
	\Psi^*(x,t) & =  \frac{1}{2\pi} \int dk \Psi^*(k,t)e^{-ikx}; \\
\Psi(y,\tau)& = \frac{1}{2\pi} \int dp \Psi(p,\tau) e^{ipy}; \\
U(x-y,t-\tau)& = \frac{1}{2\pi} \delta(t-\tau) \int dv e^{iv(x-y) }U_0|v|^\beta .
\end{align}
Thus we obtain
\begin{align}
& I(x,t)= \\ \nonumber
&  =\frac{-i}{ (2\pi)^3\hbar} \int_0^t d\tau \delta(t-\tau) \int dy  \int dv e^{iv(x-y) } U_0|v|^\beta \int dk \Psi^*(k,t)e^{-ikx} \int  dp \Psi(p,\tau)e^{ipy} -c.c.
\end{align}
Performing first the integral in $dy$ we obtain $\int dy e^{ipy} e^{-ivy}=2\pi\delta(p-v)$ and through this $\delta$-function we can integrate in $dv$, replacing everywhere $v$ with $p$. After performing also the integral in $d\tau$ we finally find
\begin{align}
I(x,t)= \frac{-i\pi}{ (2\pi)^3\hbar} \int dk \int dp \left[ \Psi^*(k,t) \Psi(p,t) |p|^\beta U_0 e^{i(p-k)x} -c.c. \right].
\label{A2}
\end{align}
Note that the integrals in $k$ and $p$ can be factorized. The two wavefunctions are known, having the explicit expression in eq.\ (\ref{A1}). The integral $\int dk \Psi^*(k,t) e^{-ikx}$ gives the wavefunction in the $x$ space, which Lenzi et al.\ have written formally in terms of the Fox function and plotted for some values of $t$. Alternatively, one could perform numerically the two transforms, for some values of $t$,
\begin{align} 
\int dk e^{-ikx}\Psi^*(k,t) , \qquad \qquad \int dp e^{ipx}\Psi(p,t)U_0 |p|^\beta 
\end{align}
and plot the imaginary part of the result. We prefer, however, to make a first order expansion in the coupling $U_0$ (see next subsection), which gives a relatively simple analytical result.

A useful check for the extra-current (\ref{A2}) is to set $t=0$, which considerably simplifies the wavefunctions in momentum space; $\Psi(k,0)$ is taken to be a Gaussian, $\Psi(k,0)=e^{-\frac{1}{2}k^2}/\sqrt{2\pi}$, whence
\begin{align}
I(x,0)=\frac{-i\pi}{ (2\pi)^3\hbar}\int dk \frac{1}{\sqrt{2\pi} } e^{-\frac{1}{2}k^2} e^{-ikx} \int dp \frac{1}{\sqrt{2\pi} } e^{-\frac{1}{2}p^2} U_0 |p|^\beta e^{ipx} - c.c.
\label{ex24}
\end{align}
The integral in $dk$ gives a real Gaussian; being interested into the imaginary part of the product, we consider only the imaginary part of the second factor, which is
\begin{align}
\frac{U_0}{\sqrt{2\pi} } \int dp [i \sin(px)] e^{-\frac{1}{2} p^2}|p|^\beta .
\label{ex25}
\end{align}
The latter integral is zero, and it follows that $I(x,0)=0$.

\subsection{First order expansion of the extra-corrent}

An analytical result for the extra-corrent (\ref{A2}) can be obtained by a first-order perturbative expansion in the coupling parameter $U_0$. Expanding the exponential in eq.\ (\ref{A1}) one finds that $\Psi(k,t)$ is given by the free wavefunction $\Psi(k,0)e^{-\frac{1}{2}tk^2}$ plus a correction of order $U_0$. Since $I(x,t)$ in (\ref{A2}) is itself proportional to $U_0$, at first order one obtains
\begin{align}
	I^{(1)}(x,t)=\frac{-i\pi U_0}{(2\pi)^3\hbar} \int dk \, e^{\frac{i}{2}tk^2} \frac{e^{-\frac{1}{2}k^2}}{\sqrt{2\pi}} e^{-ikx} \int dp \, e^{-\frac{i}{2}tp^2} \frac{e^{-\frac{1}{2}p^2}}{\sqrt{2\pi}} e^{ipx} |p|^\beta .
\end{align}

\begin{figure}
\begin{center}
\includegraphics[width=10cm,height=6cm]{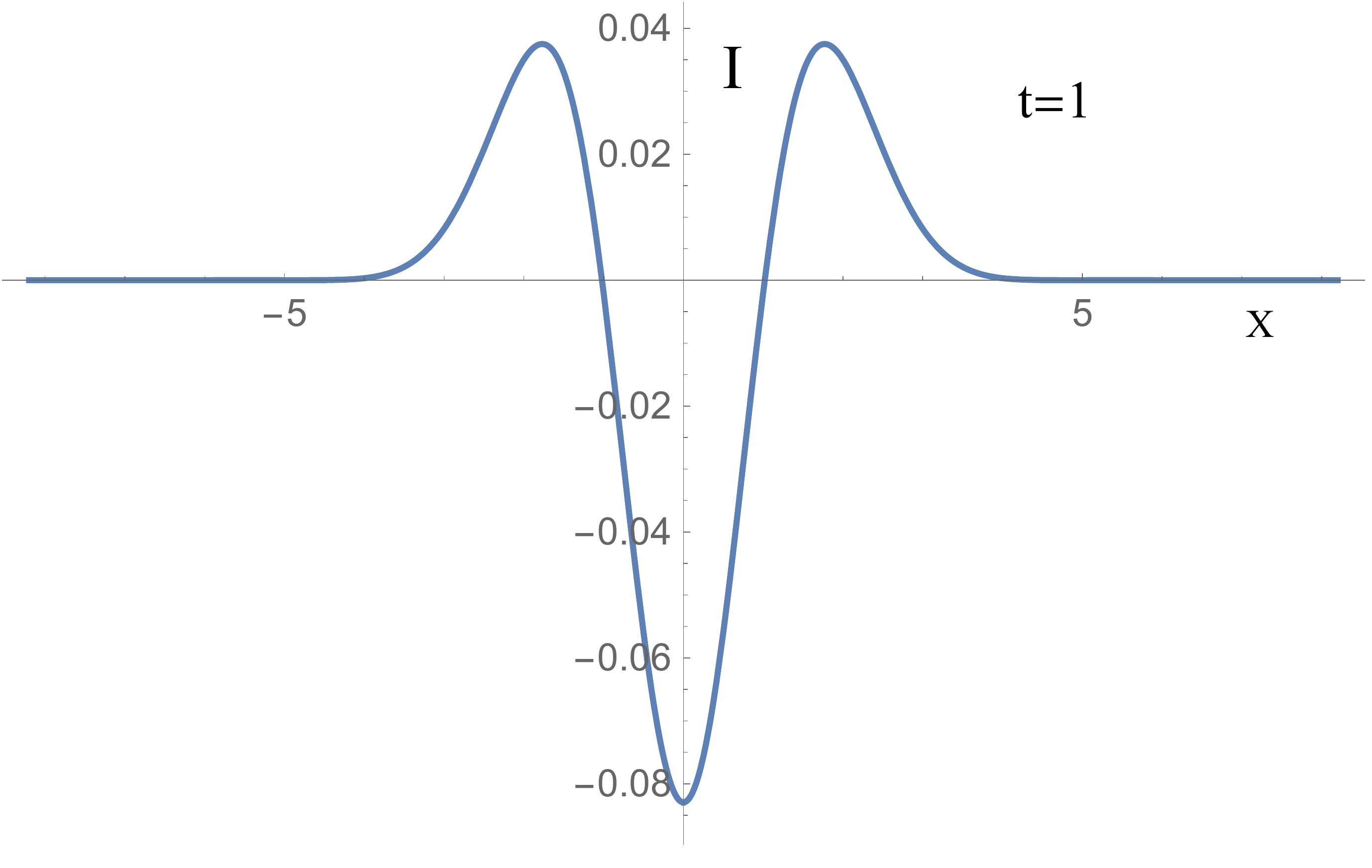}
\caption{Extra-current $I(t,x)=\partial_t \rho(t,x)+\partial_x j(t,x)$ in the spreading of a Gaussian wave packet according to the non-local equation (\ref{onda-Lenzi}). The value of $I$ is shown as a function of $x$ at the instant $t=1$. All units have been rescaled with the definition $\hbar=m=1$. The regions of negative $I$ are those where there is charge depletion (or more generally probability depletion) without a corresponding current, and conversely for the regions of positive $I$. The spreading of the wave packet is accelerated, compared to the case of locally conserved charge and $I=0$. The curves shown in these figures represent perturbative solutions of the wave equation, at first order in the coupling $U_0$. For any fixed time, $I(t,x)$ can be expressed in terms of elementary functions and Bessel functions.} 
\label{fig3a}
\end{center}  
\end{figure}

\begin{figure}
\begin{center}
\includegraphics[width=10cm,height=6cm]{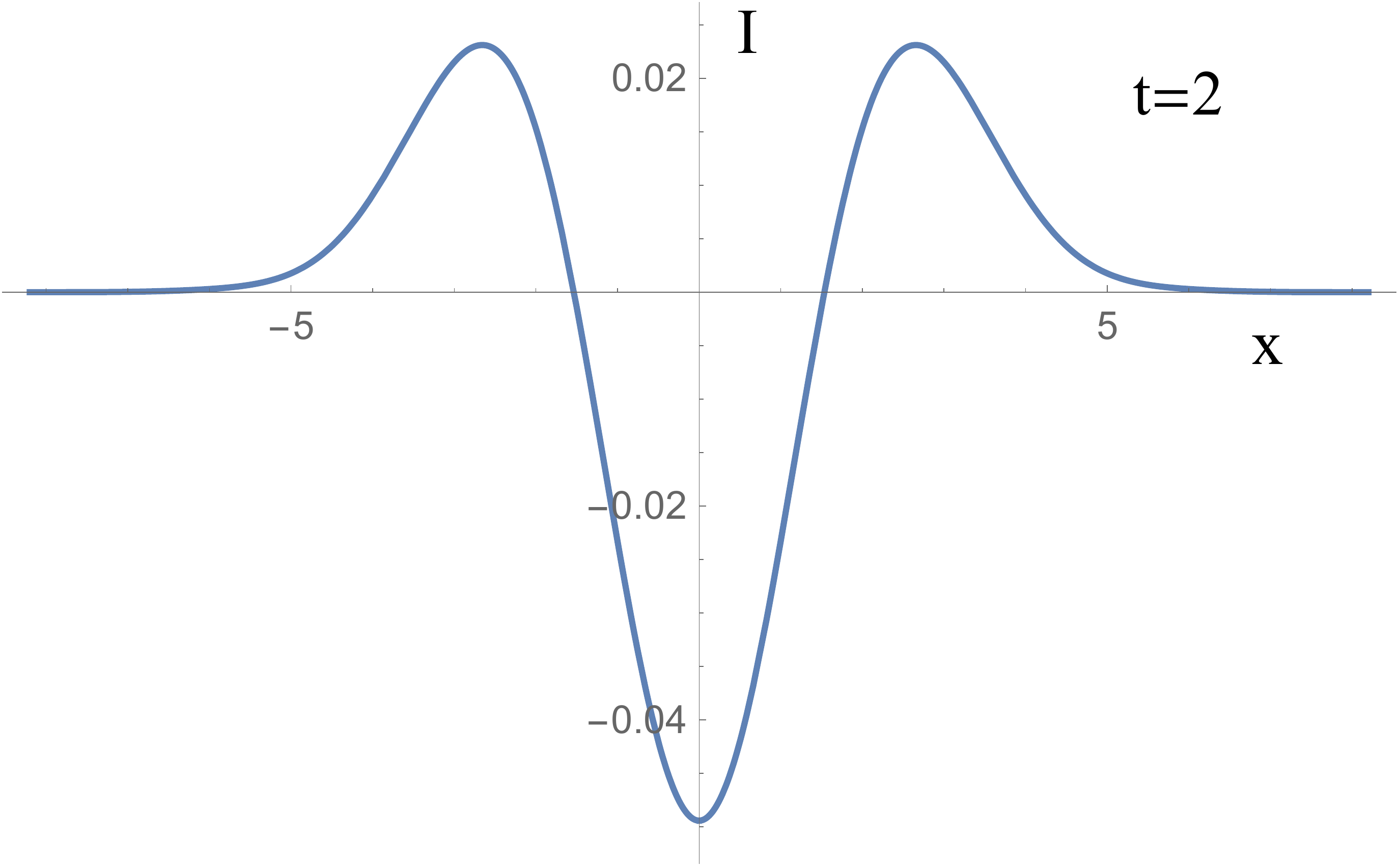}
\caption{Same as in Fig.\ \ref{fig3a}, with $t=2$.} 
\label{fig3b}
\end{center}  
\end{figure}

\begin{figure}
\begin{center}
\includegraphics[width=10cm,height=6cm]{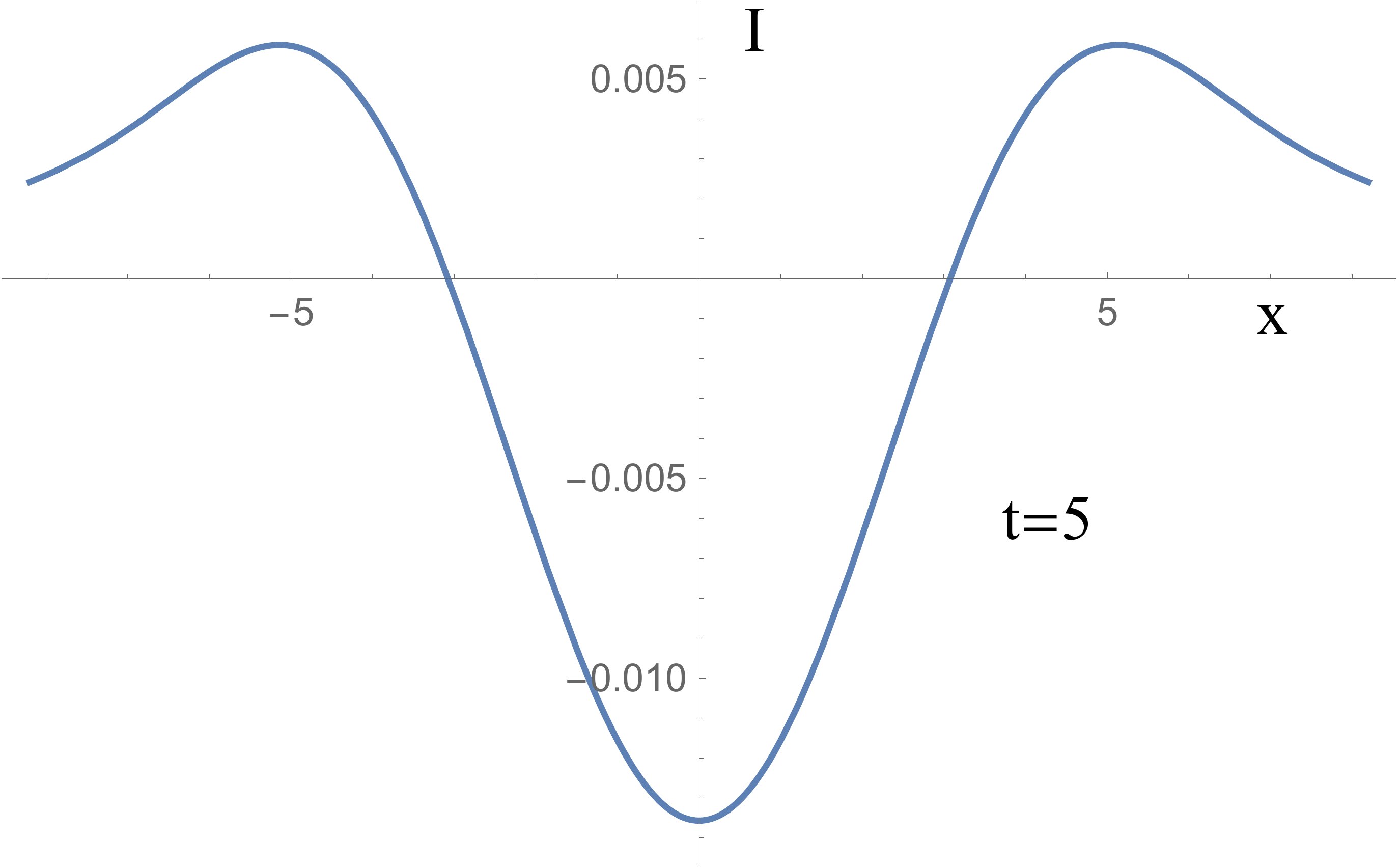}
\caption{Same as in Fig.\ \ref{fig3a}, with $t=5$.} 
\label{fig3c}
\end{center}  
\end{figure}

For fixed values of $t$ this expression can be written in terms of elementary functions and Bessel functions. In Figs.\ \ref{fig3a}, \ref{fig3b}, \ref{fig3c} the result is plotted at three succeeding times, in order to give a visual picture of the evolution. We can see that the spreading of the wave packet is accelerated, compared to the case of locally conserved charge, because some probability desappears from the center of the packet (where $I<0$) and appears at the same time near the borders (where $I>0$). This is consistent with the results of \cite{Lenzi2008solutions}, showing super-diffusion in the spreading of the packet with initial condition $\Psi(x,0)=e^{-x^2/2}/\pi^{1/4}$.

\section{Local approximation of the Gorkov equation}
\label{gorkov}

As mentioned in the Intoduction, the integral Gorkov equation (\ref{eqgorkov}) for superconductivity $F (\textbf{x})=\int d^3 y \, K(\textbf{x},\textbf{y}) F (\textbf{y})$ can be approximated under certain conditions by a differential equation of the Ginzburg-Landau type, which has a locally conserved current. We recall that \cite{waldram1996superconductivity}

\begin{enumerate}

	\item The function $F( {\bf x} )$ represents an effective wavefunction and can be written as $F( {\bf x} )=\lambda( {\bf x} ) \langle \Psi_-( {\bf x} ) \Psi_+( {\bf x} ) \rangle$, where $\Psi_\pm$ are the quantum field operators which create a particle with spin $\pm \frac{1}{2}$ at $\textbf{x}$ and $\lambda( {\bf x} )$ is a measure of the effective local attraction of the electrons, such that the two-particle potential in the Hamiltonian is $V( {\bf x},{\bf y} )=-\lambda( {\bf x} ) \delta^3( {\bf x}-{\bf y} )$.

\item The Gorkov equation can be derived from the Bogoliubov self-consistent equations, which in turn are obtained from a local quantum field theory plus a mean-field approximation.

\end{enumerate}

We summarize here the local approximation procedure given by Waldram \cite{waldram1996superconductivity,hook1973ginzburg}. Let us suppose that the order parameter $F$ varies slowly in space (this is true for low-$T_c$ superconductors, but not for high-$T_c$ superconductors, which have a very short coherence length). Expanding $F( {\bf y} )$ for ${\bf y}$ close to ${\bf x}$ we can write
\begin{equation}
 F( {\bf y} ) \simeq F( {\bf x} ) + \sum_{i=1}^3 \frac{\partial F({\bf x})}{\partial x_i} (y_i-x_i) + \frac{1}{2} \sum_{i,j=1}^3 \frac{\partial^2 F({\bf x})}{\partial x_i \partial x_j}(y_i-x_i)(y_j-x_j).
\label{taylor}
 \end{equation}
 
 Suppose that the kernel $K( {\bf x},{\bf y} )$ is spherically symmetric, i.e. depends only on $|{\bf x}-{\bf y}|$. Replacing (\ref{taylor}) into the integral equation we find the approximated local equation 
 \begin{equation}
 F( {\bf x} )\simeq AF( {\bf x} )+\frac{1}{2} S \nabla^2 F( {\bf x} )
\label{GGL}
 \end{equation}
 where
 \begin{equation}
 A=\int d^3y K( {\bf x},{\bf y} ) \qquad \qquad S=\int d^3y K( {\bf x},{\bf y} ) ( {\bf x}-{\bf y} )^2.
 \end{equation}
If $K( {\bf x},{\bf y} )$ was a really function of $|{\bf x}-{\bf y}|$ and of  nothing else, then after changing integration variable to $({\bf y}-{\bf x})$ we could conclude that the quantities $A$ and $S$ are independent from ${\bf x}$. However, in the examples of kernel given in Refs.\ \cite{waldram1996superconductivity,hook1973ginzburg}, $K$ ``has a weak dependence on $F$''. It follows that $A$ and $S$ also have a weak dependence on $F$; however, disregarding this dependence eq.\ (\ref{GGL}) reduces to a familiar wave equation of the Ginzburg-Landau type. One can then look for non-local corrections to this equation, but they turn out to be quite complicated, and little progress has been made in this direction after Ref.\ \cite{hook1973ginzburg} (for recent work see \cite{grigorishin2012nonlocal}).

The Gorkov equation has been used to analyze the proximity effect in superconductors (the diffusion of the order parameter into an adjacent normal material), especially for those cases where the microscopic BCS approach is not applicable. Both from the physical and mathematical point of view, it might be interesting to consider cases in which the kernel is independent from $F$, but cannot simply be approximated like in eq.\ (\ref{taylor}), i.e., there is a true non-locality. In that case, the form of the kernel should be postulated on phenomenological grounds. The extra-current $I(\textbf{x})$ could then be computed along similar lines as done in the previous Sections. Note that the Gorkov equation is time-independent, so the presence of an extra-current would mean that the stationary conservation law $\nabla \cdot \textbf{j}(\textbf{x})=0$ becomes $\nabla \cdot \textbf{j}(\textbf{x})=I(\textbf{x})$.

\section{Electromagnetic coupling}
\label{emc}

It is known that Maxwell equations are only compatible with a locally conserved current. In four-dimensional form, they are written as $\partial_\mu F^{\mu \nu}=4\pi j^\nu/c$, in Gauss units; since the tensor ${F^{\mu \nu }}$ is antisymmetric, it follows as necessary condition that $\partial_\mu j^\mu=0$. This is the covariant form of the continuity equation. The field equations for ${F^{\mu \nu }}$ are obtained from the Lagrangian of the pure electromagnetic field, plus a current coupling of the form ${j_\mu }{A^\mu }$, with $j^\mu=(c\rho,\textbf{j})$ and $A^\mu=(V,\textbf{A})$. In quantum mechanics, the same coupling can be introduced through the principle of gauge invariance, whereby the momentum is replaced by $\left( {{\bf{p}} - q{\bf{A}}/c} \right)$ and the energy by $\left( {E - qV} \right)$.

For quantum systems lacking a locally conserved current one can employ an extension of Maxwell equations, first introduced by Aharonov and Bohm (\cite{hively2012toward,van2001generalisation} and refs.). In \cite{Modanese2017MPLB} a new version was given, where the degree of freedom $S=\partial_\mu A^\mu$ is explicitly removed and the field equations take the form
\begin{align}
	\partial_\mu F^{\mu \nu}=\frac{4\pi}{c} \left( j^\nu+i^\nu \right);  
	\label{mme-cov}
\end{align} 
here $j^\nu$ is the usual localized physical current, of the form (\ref{cont}), and $i^\nu$ is a secondary, additional current, which is obtained applying the non-local operator $\partial^{-2}=(\partial^\alpha \partial_\alpha)^{-2}$ to the term which breaks the local conservation:
\begin{align}
i^\nu=-\partial^\nu\partial^{-2} \left(\partial_\gamma j^\gamma \right).
\label{i-sec}
\end{align}
In components, we can rewrite the secondary current as $i^0=-\partial_t \partial^{-2} \left(\partial_\gamma j^\gamma \right)$, ${\bf i}=\nabla \partial^{-2} \left(\partial_\gamma j^\gamma \right)$.
Because of the antisymmetry of $F^{\mu \nu}$, the field equations automatically imply that $\partial_\nu (j^\nu+i^\nu)=0$. The reasoning could also be reversed, starting from the latter expression to define a current $i^\nu$ which compensates the non-conservation of $j^\nu$.

We recently became aware of a work where just this compensation procedure has been carried out, for the vector current density ${\bf J}$, in order to recover local conservation in the stationary case \cite{li2008definition}. The authors of \cite{li2008definition} consider a wavefunction which obeys a wave equation with a non-local potential. They write the total current in the form ${\bf J}({\bf r})={\bf J}_c({\bf r})+{\bf J}_n({\bf r})$, where ${\bf J}_c({\bf r})$ is the analogue of our ${\bf j}$ in (\ref{cont}) and ${\bf J}_n({\bf r})$ is a non-local current density defined as ${\bf J}_n({\bf r})=-\nabla \phi({\bf r})$; in turn, $\phi({\bf r})$ is determined by solving the Poisson equation $\nabla^2 \phi({\bf r})=-\rho_n({\bf r})$, and $\rho_n({\bf r})$ is the term which breaks the continuity condition. The authors of \cite{li2008definition} regard ${\bf J}({\bf r})$ as the true physical current density and prove an important property: by integrating this current density over an appropriate surface, one obtains the Landauer-B\"uttiker formula of the total current as computed in the quantum transport theory applied to the wavefunction they consider.

In our opinion, it is not clear yet whether the total current ${\bf J}({\bf r})$ has the same physical properties of the usual ``$\rho v$ current'' of eq.\  (\ref{cont}), made with the local gradient of the wavefunction. In fact, the part ${\bf J}_n({\bf r})$ of the current ${\bf J}({\bf r})$ depends not only on the wavefunction in ${\bf r}$, but on the wavefunction in all space at the same instant (even though in the full time-dependent expression of the secondary current (\ref{i-sec}) obtained from the extended Maxwell equations it is clear that this dependence is causal and properly retarded). All this can have non-standard implications also for dissipation, in the case when ${\bf J}_n({\bf r})$ ``flows'' at a different place than ${\bf J}_c({\bf r})$. 

{We see here an example of a more general problem concerning non-local theories, namely: what is the correct physical interpretation of the wavefunction, if we admit that the usual interpretation rule of quantum mechanics (Born rule) is only justified by a local conservation law?}

In any case, from all the work cited above it can be safely concluded that, at least for the wavefunctions considered in \cite{li2008definition}, 

(1) Conduction in the presence of non-locally-conserved currents satisfies the Landauer-B\"uttiker formula of quantum transport.

(2) The electromagnetic field generated by these currents is given by the extended Maxwell equations (\ref{mme-cov}).

In the three-dimensional vector formalism the extended equations without sources maintain the usual form, namely $\nabla \times \textbf{E}=-(1/c)(\partial \textbf{B}/\partial t)$, $\nabla \cdot \textbf{B}=0$. The extended equations with sources are written as follows \cite{modanese2017electromagnetic}
\begin{align}
	& \nabla \cdot \textbf{E}=4\pi \rho-\frac{1}{c^2}\frac{\partial}{\partial t}\int d^3y \frac{I\left(t_{ret},\textbf{y} \right)}{\left|\textbf{x}-\textbf{y} \right|}; \label{eqE} \\
	& \nabla \times \textbf{B}-\frac{1}{c} \frac{\partial \textbf{E}}{\partial t}=\frac{4\pi}{c} \textbf{J}+\frac{1}{c} \nabla \int d^3y \frac{I\left(t_{ret},\textbf{y} \right)}{\left|\textbf{x}-\textbf{y} \right|}, \label{eqB}
\end{align}
where $I$ denotes the scalar quantity which breaks local charge conservation:
\begin{align}
	I=\left(\partial_\gamma J^\gamma \right)=\frac{\partial\rho}{\partial t}+\nabla \cdot \bf{J}.
\end{align}

The extended Maxwell equations have an important ``censorship'' property  \cite{Modanese2017MPLB}: they imply that measurements of the field strength made with test particles cannot reveal any possible local non-conservation of the source. The apparent conservation is due to the secondary source, which however may be extended in space, outside the physical source. See our comments on this point in the Introduction. 

Some solutions of the equations (\ref{eqE}), (\ref{eqB}) have been given in \cite{Modanese2017MPLB,modanese2017electromagnetic} and others will be presented in a forthcoming work. {The explicit expressions for the extra-current found in this work serve as guidance for non-local quantum sources for these equations. More realistic sources could be obtained, as mentioned, from the Gorkov equation.}

We finally note that changes in quantum electrodynamics following an extension of Maxwell theory with the Aharonov-Bohm Lagrangian have been extensively studied by Jimenez and Maroto (\cite{jimenez2011cosmological} and refs.). However, they only consider coupling of the electromagnetic field to locally conserved currents. As a consequence, additional degrees of freedom like longitudinal wave components are decoupled from matter, but can have cosmological relevance.

\section{Conclusions}
\label{conc}

The non-locality of the fractional Schr\"odinger equation and of the Schr\"odinger equation with a non-local interaction potential leads to a violation of the local conservation law of probability density (and of charge, in the case when the wave equation describes a charged particle). We have demonstrated this explicitly both for 1D wave packets in motion and at rest, interpreting physically the extra-current $I(t,x)=\partial_t \rho(t,x)+\partial_x j(t,x)$ as the consequence of a kind of super- or sub-diffusion, depending on the cases. We have checked that at the microscopic level the magnitude order of the extra-current $I(t,x)$ is comparable to that of the current gradient $\partial_x j(t,x)$, so that the effect cannot be disregarded as a minor correction. We have discussed the impact of these properties on the electromagnetic coupling of the wave functions considered, recalling the ``censorship'' property of extended electrodynamics, which to some extent prevents the non-locality of the wavefunctions from causing a non-local behavior of the electromagnetic field. A brief discussion has been given of a possible application of these ideas to the Gorkov integral equation for the proximity effect in superconductors.

{We recall that non-local field theories have been discussed by several authors, starting from early works by W.\ Pauli and others. The subject is wide and complex, also because non-locality can occur in different ways and accordingly it can have different mathematical realizations. Substantial progress has recently been made by Kegeles and Oriti \cite{kegeles2016generalized}. They extend N\"other's theorem to non-local field theories, thus finding a more general relation between symmetries of the action and conserved quantities. 
They consider as non-local an action functional defined by a Lagrangian
which mathematically depends on multiple copies of a jet bundle. Theories described by this kind of actions are employed for many-particle systems in various branches of physics such as hydrodynamics, condensed matter and solid state physics. An example is given by effective theories in which one disregards some microscopic degrees of freedom; the price to pay is often the emergence of non-locality. 
The general results of \cite{kegeles2016generalized} are applied to a model of a complex scalar field with a non-local two-body interaction, appearing in the theory of Bose-Einstein condensates.
A comparison of the
currents of the local and non-local theories explicitly shows the appearance of a non-local correction term.}

{The scope of our work has been more limited. We have considered certain non-local wave equations as effective physical models, without postulating an underlying action. We have found some solutions that can be used as prototype sources for extended Maxwell equations (which also automatically generate non-local corrections terms, while preserving the local covariant Lagrangian coupling $j^\mu(x) A_\mu(x)$). Results of the extended Maxwell equations, including a proposal of experimental verification, will be presented in a forthcoming work.
}

\bigskip
\noindent
\textbf{Acknowledgments:} We would like to thank an anonimous reviewer for pointing out recent important applications of the fractional Schr\"odinger equation to optics (Sect.\ 1), and another reviewer for bringing to our attention Refs.\ \cite{li2008definition,kegeles2016generalized} and their implications.

\bibliographystyle{unsrt}
\bibliography{mme}

\end{document}